\documentclass{PoS}

\newcommand\ttiny[1]{\textrm{\tiny{#1}}}
\newcommand{\beq}{\begin{equation}}
\newcommand{\eeq}{\end{equation}}

\title{A new pQCD based Monte Carlo event generator for jets in the quark-gluon plasma}

\ShortTitle{A new Monte Carlo event generator for jets in the QGP}

\author{\speaker{Paul Caucal}\\
        Institut de Physique Th\'{e}orique, Universit\'{e} Paris-Saclay, CNRS, CEA, F-91191, Gif-sur-Yvette, France\\
        E-mail: \email{paul.caucal@ipht.fr}}

\author{Edmond Iancu\\
       Institut de Physique Th\'{e}orique, Universit\'{e} Paris-Saclay, CNRS, CEA, F-91191, Gif-sur-Yvette, France\\
       E-mail: \email{edmond.iancu@ipht.fr}}
\author{Alfred H. Mueller\\
    Department of Physics, Columbia University, New York, NY 10027, USA\\
   E-mail: \email{amh@phys.columbia.edu} }
       
 \author{Gregory Soyez\\
       Institut de Physique Th\'{e}orique, Universit\'{e} Paris-Saclay, CNRS, CEA, F-91191, Gif-sur-Yvette, France\\
       E-mail: \email{gregory.soyez@ipht.fr}}
       
\abstract{A main difficulty in understanding the dynamics of jets produced in the high-density environment of ultrarelativistic heavy ion collision, 
is to provide a unified description for the two sources of radiation that are a priori expected: the "vacuum-like" emissions responsible for the parton 
shower from large virtualities down to the hadronisation scale and the "medium-induced" emissions responsible for the energy 
loss by the jet. In the recent paper \cite{Caucal:2018dla}, we demonstrated that these two mechanisms can be factorized from each other within a controlled, 
"double-logarithmic" approximation in perturbative QCD. In this proceeding we recall the main features of the jet evolution in a dense QCD medium. We 
emphasize that the in-medium parton showers differ from those in the vacuum in two crucial aspects: their phase-space is reduced and the first emission outside the medium can violate angular ordering. 
Based on this factorized picture, which is Markovian, we have recently developed a Monte Carlo event generator which includes both vacuum-like and medium-induced emissions and goes beyond the double logarithmic 
approximation. We here present our first results for the fragmentation function and the energy loss by the jet.}

\FullConference{International Conference on Hard and Electromagnetic Probes of High-Energy Nuclear Collisions\\
		30 September - 5 October 2018\\
		Aix-Les-Bains, Savoie, France}

\begin{document}

\section{Introduction}

One important goal of the experimental programs at RHIC and at the LHC is the characterisation of the quark-gluon plasma (QGP) produced in
ultrarelativistic heavy ion collisions. Jet quenching observables,  i.e.  the modifications of the properties of an energetic jet due to its interactions with the
dense form of QCD matter produced right after a collision, are particularly interesting to study the properties of this deconfined phase.

The suppression of the jet cross-section in nucleus-nucleus collision w.r.t proton-proton 
was the first historical indication of jet energy loss in the plasma and has been accurately measured at both RHIC and the LHC  \cite{Adler:2002xw,Adcox:2001jp,Khachatryan:2016jfl,Aaboud:2018twu}. Most recently, 
substructure observables such that the jet fragmentation function \cite{Chatrchyan:2014ava,Aaboud:2018hpb}
have provided new 
insights on the evolution of a jet in the QGP. 

From a theoretical point of view, high-$p_T$ jets are peculiarly advantageous because one can rely on perturbative QCD (pQCD) to understand their properties. Within this approach, the dense 
weakly-coupled quark gluon medium may trigger medium-induced radiations from the virtual partons inside jets.
This can be
computed with the BDMPS-Z formalism \cite{Baier:1996kr,Zakharov:1996fv,Wiedemann:2000za}, recently extended to include multiple branchings 
\cite{Blaizot:2012fh,Blaizot:2013hx,Blaizot:2014ula,Kurkela:2014tla,Escobedo:2016jbm}. 
Nevertheless, the global jet structure has to be built with the usual, ``vacuum-like'' bremsstrahlung process through which a parton loses its virtuality from the hard scale until the hadronisation.

In the recent paper \cite{Caucal:2018dla}, a landmark has been achieved in the understanding of the interplay between these two mechanisms, vacuum-like emissions (VLEs) and medium induced radiations. 
We showed that the double-logarithmic approximation (DLA) of pQCD provides a rigorous framework to deal with the jet evolution in a QGP  and should be regarded as a first step toward a 
more advanced treatment including other physical effects.

In this proceeding, we first recall the crucial aspects of double-logarithmic parton cascades in the QGP. Then, we explain how the probabilistic picture that emerges from the DLA
has been implemented in a new Monte Carlo (MC) event generator which accounts for both vacuum-like emissions and medium induced radiations. We finally present preliminary results 
for the nuclear modification factor and 
the jet fragmentation function.

\section{Jet evolution in the presence of a medium at DLA}

After having recalled the basic properties of jet evolution in the vacuum, this section sums up how the medium changes the development of parton cascades.

\subsection{Evolution in the vacuum}

In the vacuum, the two building blocks to construct a parton shower are the \textit{Bremsstrahlung process} which provides the probability law for a soft and collinear emission by a virtual parton and 
the \textit{angular ordering} property which enables one
to treat successive Bremsstrahlung emissions as a Markov chain of independent elementary radiation events, ordered in angles \cite{Dokshitzer:1991wu}.

The specific feature of the Bremsstrahlung probability distribution $d^2\mathcal{P}_{\mathcal{B}}$ is the logarithmic enhancement of soft and collinear emissions: 
\beq d^2\mathcal{P}_{\mathcal{B}}=\frac{\alpha_s C_R}{\pi}\frac{d\omega}{\omega}\frac{d\theta^2}{\theta^2} \eeq
The essence of the DLA is the resummation of contributions coming from such processes to all orders in the computation of intrajet
observables.

\subsection{Factorization between VLEs and medium induced radiations}

In addition to that, the dense weakly-coupled medium triggers radiations \`a la BDMPS-Z if the formation time of the medium-induced parton is much larger than the mean free path between successive collisions off 
the medium constituents. Thereafter,
our QGP is characterized by only two parameters: the distance $L$ covered by the jet inside the medium and the jet quenching parameter $\hat{q}$ corresponding to the averaged 
transverse momentum acquired by multiple collisions during a time $\Delta t$ via the relation $\langle k_\perp^2\rangle=\hat{q}\Delta t$.

In this framework, the probability distribution 
$d\mathcal{P}_{\ttiny{med}}$ for medium induced processes is well approximated by the formula \cite{Salgado:2003gb,Wiedemann:2000tf} 
\beq d\mathcal{P}_{\ttiny{med}}\simeq \frac{\alpha_sN_c}{\pi}L \sqrt{\frac{\hat{q}}{\omega^3}} d\omega \eeq
This spectrum shows no collinear nor soft logarithmic enhancement so one can safely ignore the medium-induced radiations when computing the intrajet multiplicity at DLA.

\begin{figure}
 
 \begin{center}
  \begin{tabular}{cc}
 \includegraphics[width=70mm]{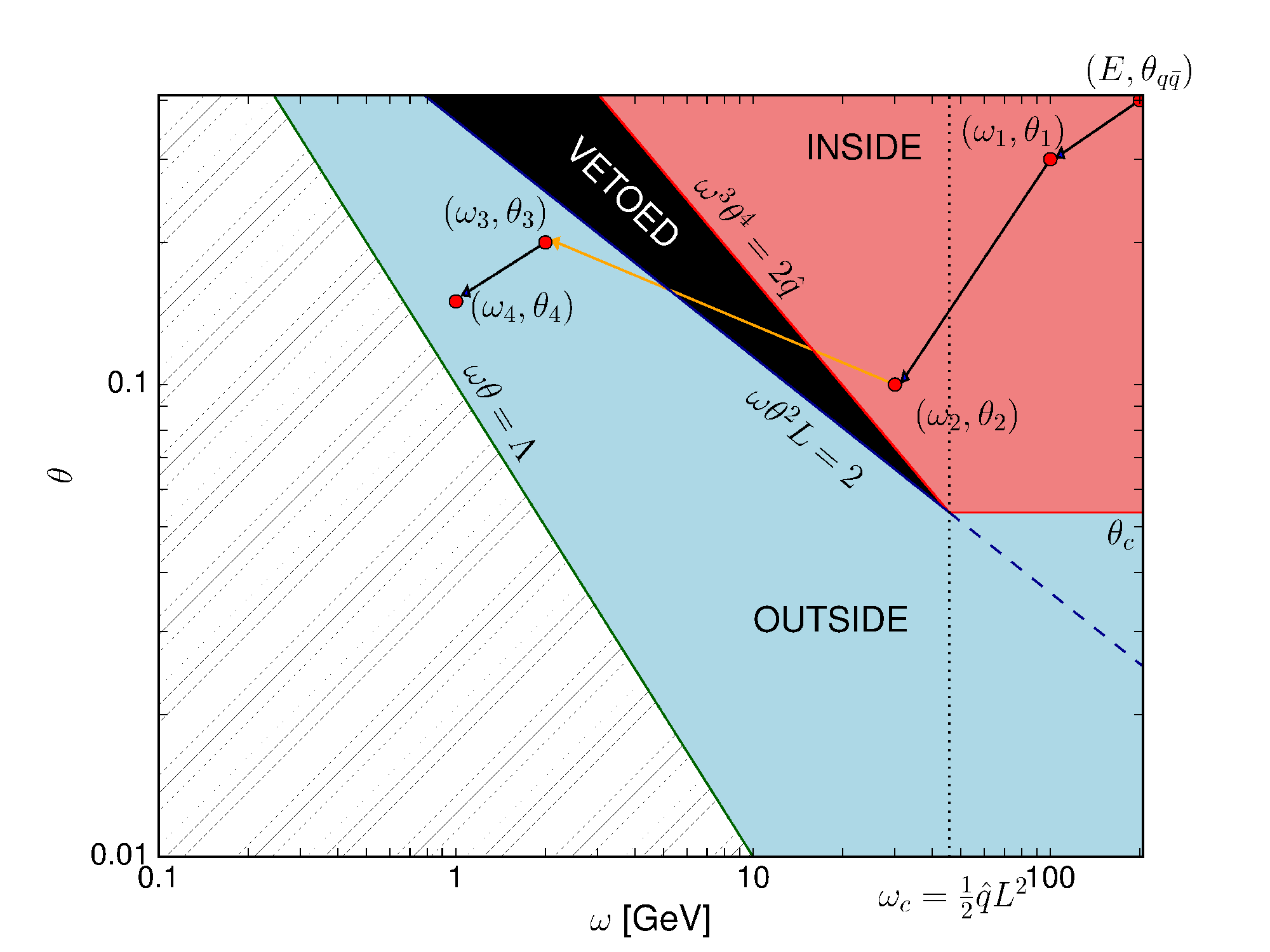}&
 \includegraphics[width=70mm]{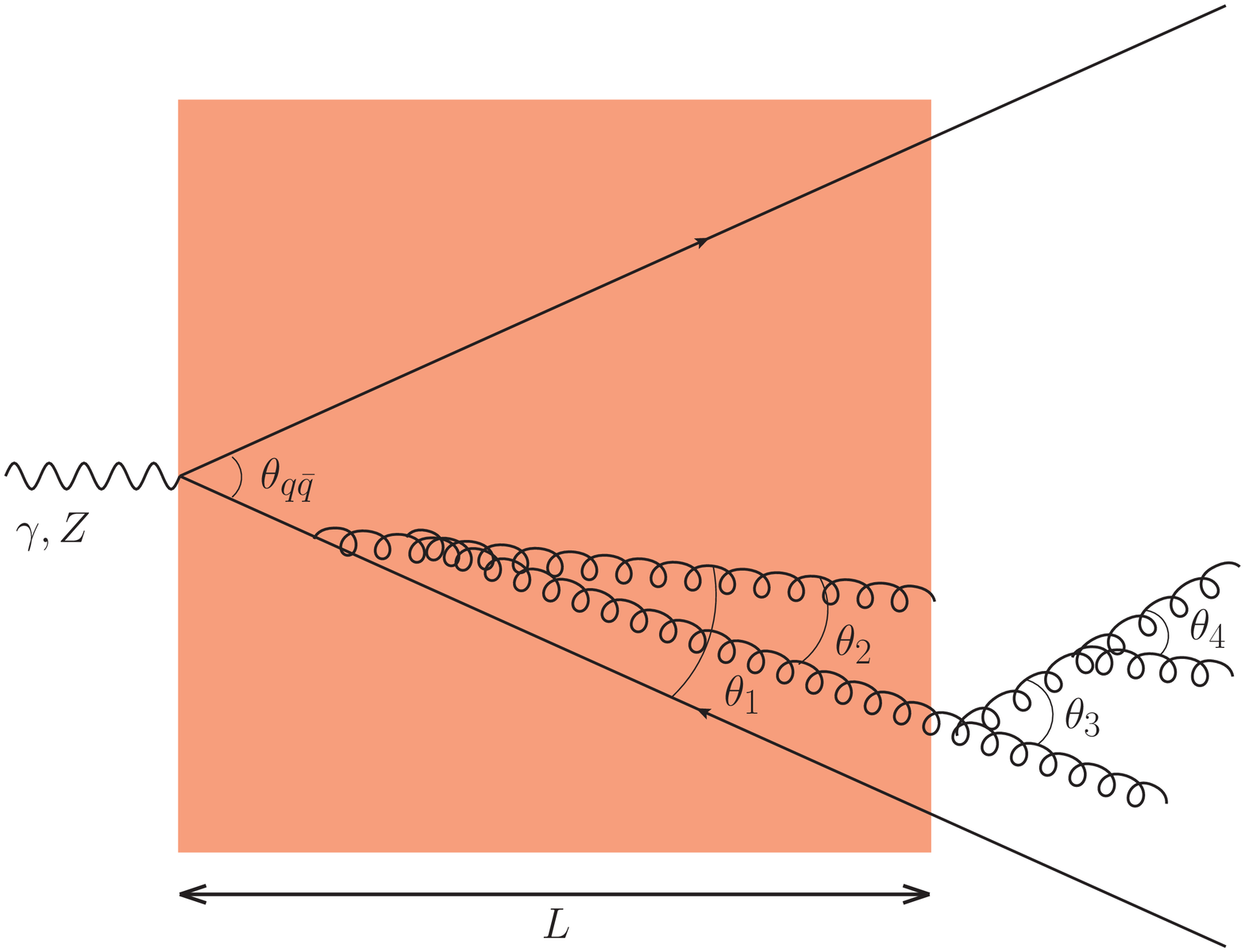}
 \end{tabular}
 \end{center}
 \caption{\label{phasespace}(Left) Phase space for one VLE in a jet of energy $E=200$ GeV and opening angle $\bar{\theta}=0.4$ in the presence of a medium. Above the red line, $k_\perp\ge k_{\perp}^{\ttiny{min}}=\sqrt{\hat{q}t_f}$ and 
 below the blue line $t_f\ge L$. In the hatched region $k_\perp\le\Lambda_{\ttiny{QCD}}$, partons have hadronised and the pQCD regime is not valid anymore. 
 (Right) Cascade which might happen
 with a medium but not in the vacuum. The position of corresponding successive emissions in the phase space are represented on the left. Angular ordering is violated by the third emission, but preserved by all the others. The yellow arrow 
 between the second and the third  emission represents all the medium induced processes that can affect the parton 2 traveling through the medium over a large distance.}
\end{figure}

Even if at DLA, the medium seems to have no direct effects, it will appear has a new constraint in the phase space for VLEs and this constraint will lead to the factorization between the vacuum-like showers and the medium 
induced effects. Indeed, a VLE occurring \textit{inside} a medium have a minimal transverse momemtum $k_{\perp}^{\ttiny{min}}=\sqrt{\hat{q}t_f}$ given by the momentum acquired via multiple collisions during its formation time $t_f$.
This minimal transverse momentum condition, 
once translated in terms of the angle and energy of emission through the uncertainty principle relation $t_f\simeq1/\omega\theta^2$ becomes a lower bound in the phase space represented by the line $\omega^3\theta^4=2\hat{q}$ in figure \ref{phasespace}-left. Of course, since the medium has a finite length, a 
VLE can also take place directly \textit{outside} the medium if its formation time $t_f$ is larger than $L$. Again, this constraint on VLE \textit{outside} the medium translates into an upper bound in the 
energy/angle phase space represented by the line $\omega\theta^2L=2$ in figure \ref{phasespace}-left.

As an immediate consequence of the previous discussion, the phase space for VLEs is reduced by the presence of a vetoed region where there is no VLE permitted as shown in figure \ref{phasespace}-left. 

As already outlined, another important consequence is the possibility to \textit{factorize} VLEs from medium induced emissions: the latter are neglected at DLA but 
we would eventually like to include them to account for energy loss and also to compute the jet fragmentation beyond DLA. The probabilistic picture for the jet evolution that emerges from the DLA analysis is the following: firstly, a vacuum-like shower 
develops itself during a very short time $\ll L$ in the in-medium region (``inside'' region, figure \ref{phasespace}-left). Then the partons that come out of this shower undergo medium-induced processes. Especially, they become seeds for 
medium-induced mini-jets as studied in references \cite{Blaizot:2013hx,Blaizot:2014ula,Kurkela:2014tla,Escobedo:2016jbm,Fister2015}. Then, when the partons come out of the medium (they are in the ``outside'' region, figure \ref{phasespace}-left), an usual angular ordered vacuum shower decreases their 
virtualities down to the hadronisation scale $\Lambda_{\ttiny{QCD}}$. 


\subsection{Violation of angular ordering}

Medium induced radiation is not the only effect of the QGP that matters for the jet evolution. When previously discussing parton showers in the vacuum, we outlined the importance of the color coherence and angular ordering to describe 
a jet as a branching process. 

A color singlet dipole with opening angle $\bar{\theta}$
propagating through the medium may lose its color coherence via multiple scattering off the medium: the two prongs suffer independent color rotations, hence the probability that the dipole remains in a color singlet
state decreases with time \cite{MehtarTani:2010ma,CasalderreySolana:2011rz,CasalderreySolana:2012ef}. The two legs of the antenna start behaving like independent color sources after a
time $t_{coh}(\bar{\theta})\equiv(4/\hat{q}\bar{\theta}^2)^{1/3}$ \cite{CasalderreySolana:2011rz}. Consequently, angular ordering could be violated in the presence of a medium.

Yet, in Ref \cite{Caucal:2018dla}, we argued that this is not the case for VLEs occurring inside the medium: the respective formation times are small enough
for the effect of color decoherence to be negligible. Hence, to DLA at least, angular ordered cascades develop inside the medium exactly as in the vacuum, as drawn in
figure \ref{phasespace}-right.

However, the last antenna produced inside the medium by the in-medium vacuum-like shower plays a special role. Indeed, this antenna will have to travel through the medium over a distance of order $L$. Calling $\theta_{last}$ 
its opening angle, two cases are possible: either $t_{coh}(\theta_{last})\le L$ or $t_{coh}(\theta_{last})>L$. In the first case which corresponds to $\theta_{last}\ge\theta_c\equiv 2/\sqrt{\hat{q}L^3}$ the last antenna hast lost its color coherence so the next emission outside the medium can radiate at any angle. 
In the second case, the antenna is still coherent at the time of formation of the next one. Consequently, the angle of the next emission is constrained by the angle $\theta_{last}$. The introduction 
of the critical angle $\theta_c$ leads to a  modification of the ``inside'' region in figure \ref{phasespace}-left: this region is now defined by $k_\perp\ge\sqrt{\hat{q}t_f}$ and 
$\theta\ge\theta_c$. In this region, cascades are always angular ordered.

To summarize, the DLA analysis of jet evolution in a dense QCD medium has exhibited three important features: a vetoed region for VLEs, the factorization of medium-induced processes between the angular ordered in-medium  and out-medium 
showers and finally one violation of angular ordering by the first emission
outside the medium is permitted \cite{Caucal:2018dla,Mehtar-Tani:2014yea}.

\section{Monte Carlo event generator}

At this stage of the discussion, it should be clear that the probabilistic picture sketched in the previous section is particularly suitable for a MC implementation which presents also the advantage to take easily into account 
the conservation of energy and the running of the strong coupling constant through the evolution.

\subsection{Brief description of the Monte Carlo implementation}

The MC event generator uses two modules: the first one deals with the vacuum-like angular-ordered shower (inside or outside the medium) and the second 
one implements the medium-induced cascades for which the branching rate is given by the BDMPS-Z rate following the ideas developed in Refs \cite{Blaizot:2013hx,Blaizot:2014ula}.

\subsubsection{Vacuum-like shower}

In order to generalize the results obtained in Ref \cite{Caucal:2018dla} for the fragmentation function 
at DLA, the vacuum-like shower implements the energy conservation and DGLAP splitting probabilities at every branching.
The color representations of the partons (either quark $q$ or gluons $g$) are properly treated.

However at this 
level of accuracy, the angular ordering property is valid only after averaging over the azimuthal angle of emission. 
To be consistent with this principle from pQCD, an azimuthal angle is chosen randomly between 0 and $2\pi$ at each emission.

Because of angular ordering, the evolution parameter of the branching process is naturally the angle of a parton 
with respect to its emitter. In the vacuum, the Sudakov form factor $\Delta_i(\bar{\theta}^2,\theta_0^2)$ is the probability to have no branching between the angle $\bar{\theta}$ of the parent parton of type $i\in\{q,g\}$ with energy $E$ and a smaller angle $\theta_0$:
\beq \label{sudavac} \Delta_i(\bar{\theta}^2,\theta_0^2)=\exp\Big(-\int_{\theta_0^2}^{\bar{\theta}^2}\frac{d\theta^2}{\theta^2}\int_{0}^{1} dz \frac{\alpha_s(zE\theta)}{2\pi}P_{ji}(z)\Theta(zE\theta>k_\perp^c)\Big) \eeq
where $P_{ij}(z)$ are the leading order DGLAG splitting functions and $\alpha_s(zE\theta)$ is the one-loop QCD running coupling evaluated at the transverse momentum of the emitted parton.

The step function $\Theta(zE\theta>k_\perp^c)$ enforces the condition that the $k_\perp\simeq zE\theta$ of the emission has to be larger than a given cut-off $k_\perp^c$. 
Inside the medium, this $k_\perp^c$ corresponds to the boundary of the ``inside'' region in figure \ref{phasespace}-left while outside the medium, this is a scale of order $\Lambda{\ttiny{QCD}}$.
In practice, a Sudakov veto algorithm based on \ref{sudavac} is used to produce a branching angle.

\begin{figure}
 
 \begin{center}
  \begin{tabular}{cc}
    \includegraphics[width=70mm]{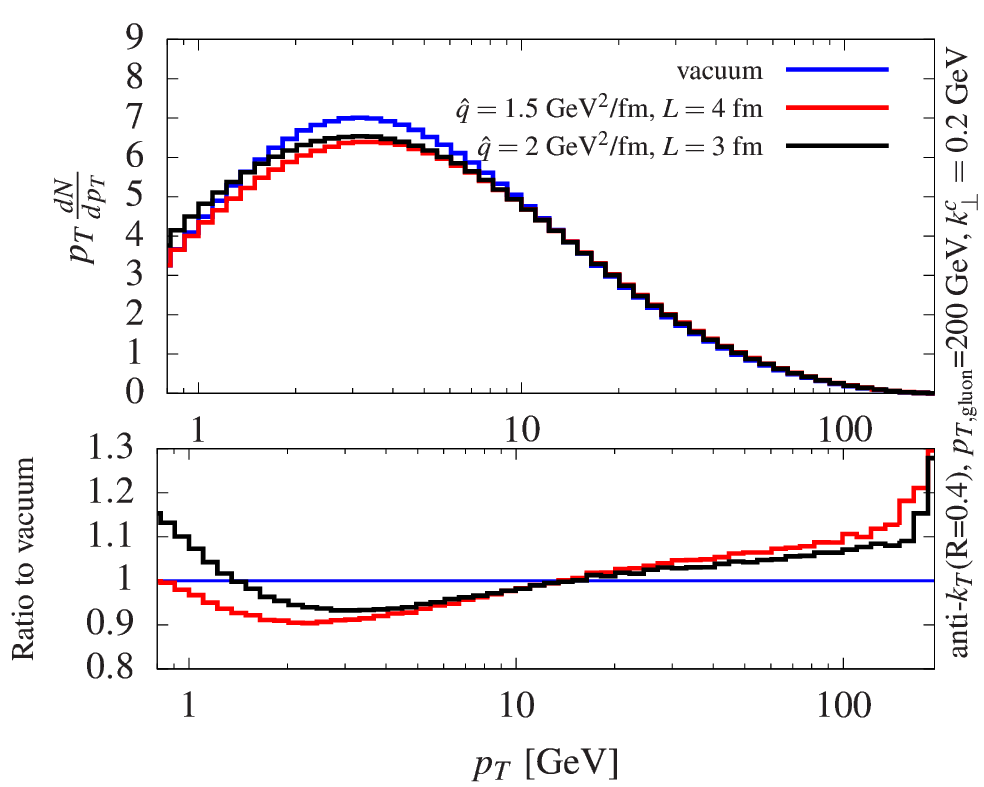}&
    \includegraphics[width=70mm]{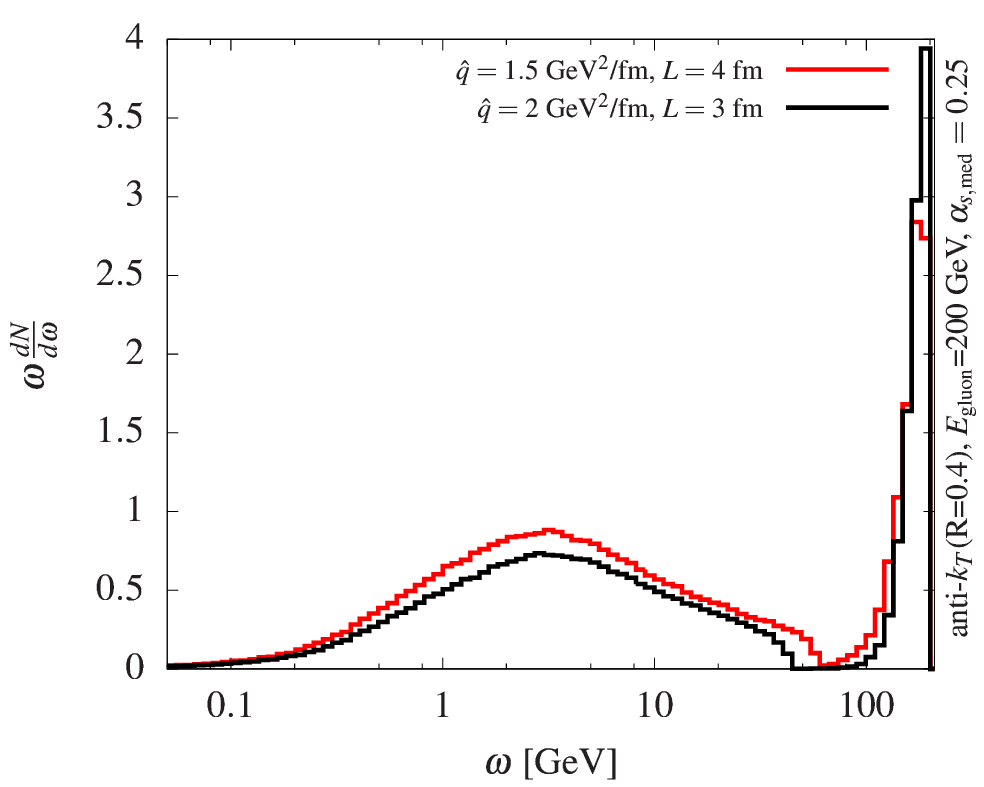}
  \end{tabular}
 \end{center}
\caption{\label{plotmodules}(Left) In blue, the fragmentation function for a jet in the vacuum created by a gluon with $p_{T,\ttiny{gluon}}=200$ GeV  as given by the module 
``vacuum-like shower'' of our MC event generator. The other curves correspond to the fragmentation function in the presence of a medium with \textit{only} VLEs for two sets of medium parameters. Below, the ratios with respect to the vacuum are plotted. (Right) The 
energy spectrum of particles produced by medium-induced splittings from a leading gluon with $E_{\ttiny{gluon}}=200$ GeV resulting from the module ``medium-induced shower'', for two 
different values of $\hat{q}$ and $L$. The parameters $z_c$ and $\theta_{\ttiny{max}}$ are fixed respectively to $10^{-5}$ and 1.}
\end{figure}

In figure \ref{plotmodules}-left, we show our MC results for the fragmentation function as produced by VLEs alone. Accordingly, the only effect of the medium is to introduce the vetoed region and the reopening of the phase space for the 
first emission outside. For comparison, we also show our MC predictions for jets propagating in the vacuum as well as the medium/vacuum ratio. The enhancement visible in this ratio at 
low $p_T$ is a consequence of decoherence whereas the depletion at intermediate energy is a consequence of the veto. There is also an enhancement at large $p_T$ 
due to the fact that, in the presence of the vetoed region, the leading particle has less probability to radiate and hence a larger probability to carry a large value of $z$.

\subsubsection{Medium-induced parton shower}

This module develops a medium-induced parton shower from a leading parton with energy $E$. The evolution parameter in the case of medium induced cascades is the ``time'' $t=x^+$ in light-cone coordinates
 with the longitudinal axis defined by the direction of motion of the leading particle \cite{Blaizot:2013hx}.
The branching rate associated with the splitting of a gluon with energy fraction $x=\omega/E$ into two gluons, one carrying away a fraction $z$ of the energy is \cite{Blaizot:2013hx}
\beq \frac{d^2\mathcal{P}_{br}}{dz dt}=\frac{1}{2\pi}\frac{P_{gg}(z)}{\tau_{br}(z,x)}\textrm{ , }\tau_{br}(z)=\frac{1}{\alpha_s}\sqrt{\frac{z(1-z)xE}{\hat{q}_{\textrm{\tiny eff}}(z)}}\textrm{ , }\hat{q}_{\textrm{\tiny eff}}(z)=\hat{q}(1-z(1-z)) \eeq
A similar formula exists for a quark splitting and is incorporated as well in our code. In the medium-induced parton shower, the strong coupling constant 
is assumed to take a fixed value $\alpha_{s,\textrm{\tiny med}}$. The treatment of the running of the coupling is left for further studies.

To produce a branching, one needs first to compute the Sudakov form factor associated with this branching rate in order to generate the next branching time.
The probability $\Delta_{\textrm{\tiny med}}(\bar{t},t_0)$ to have no branching betweem times $\bar{t}$ and $t_0$ is 
\beq \Delta_{\textrm{\tiny med}}(\bar{t},t_0)=\exp\Big(-\int_{\bar{t}}^{t_0}dt\int_{z_c/x}^{1-z_c/x} dz \frac{d^2\mathcal{P}_{br}}{dz dt}\Big) \eeq
with a cut-off $z_c$ in the energy fraction.

Here again, a Sudakov veto algorithm is used to pick the subsequent branching time.  Then, we generate a given energy fraction $z$ according to the effective splitting probability $P_{gg}(z)/\tau_{br}(z,x)$. 

At this stage, every splitting is assumed to be collinear with the leading particle. 
Indeed, as well known most of the transverse momemtum broadening is acquired during the propagation time $\Delta t$ between 
two successive branchings. In our MC, this is mimicked by ascribing to each parton an average transverse momentum 
$\langle k_\perp^2\rangle=\hat{q}\Delta t$ with a Gaussian distribution.
Evolution is stopped if the branching time is 
larger than the length of the medium $L$ or if the momentum broadening has pushed the angle of emission beyond a maximal angle $\theta_{\ttiny{max}}\simeq 1$ or finally if the energy 
fraction with respect to the leading particle carried by the parton is smaller than $z_c$.

In figure \ref{plotmodules}-right, we show the resulting energy spectrum of the medium-induced shower from an initial gluon with $E=200$ GeV. The large narrow peak near $\omega=200$ GeV is the remnant of the leading particle. 
The second broad peak corresponds to the accumulation around the characteristic transverse momemtum scale $Q_s=\sqrt{\hat{q} L}$ of soft gluons produced via medium induced 
radiations due to momentum broadening. This spectrum agrees with previous studies \cite{Blaizot:2013hx,Blaizot:2014ula,Escobedo:2016jbm,Fister2015}.

\begin{figure}

 \begin{center}
  \begin{tabular}{cccc}
    \includegraphics[width=57mm]{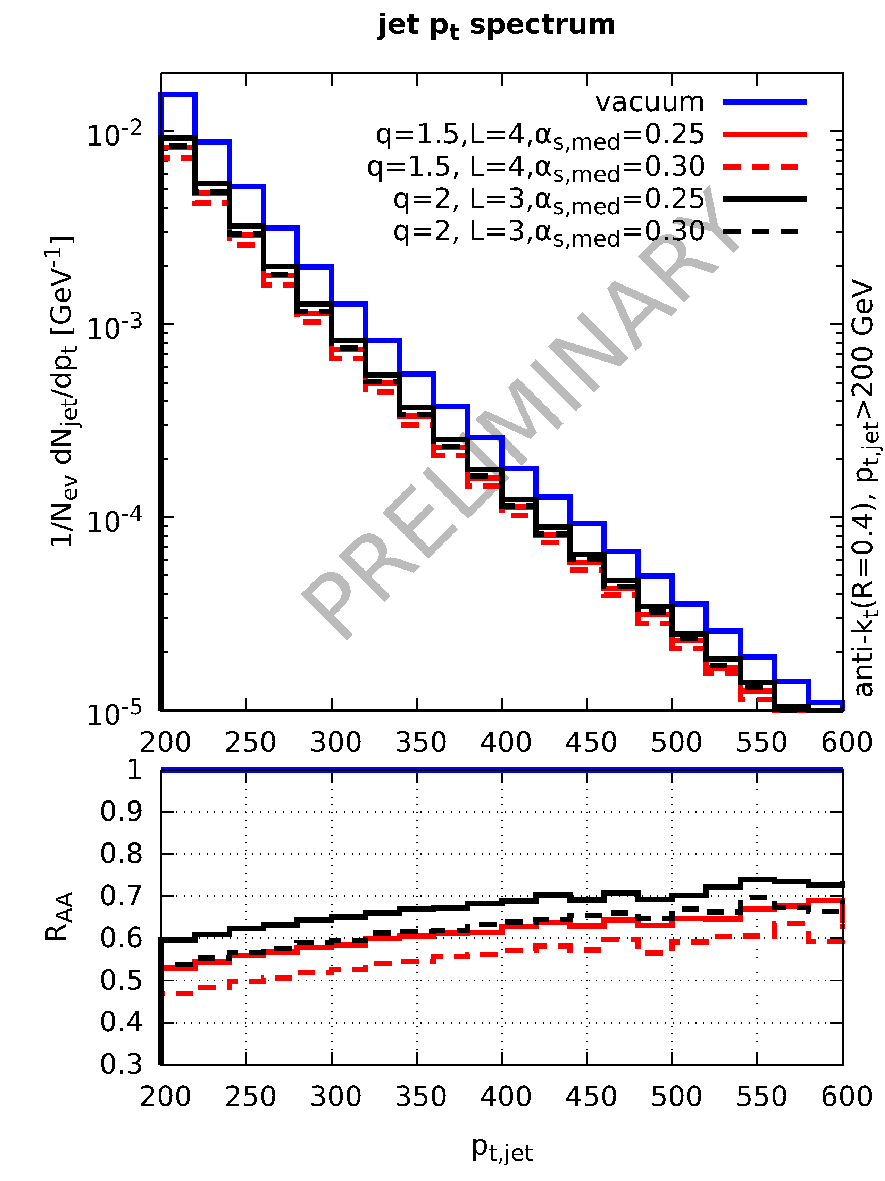}& &&
    \includegraphics[width=57mm]{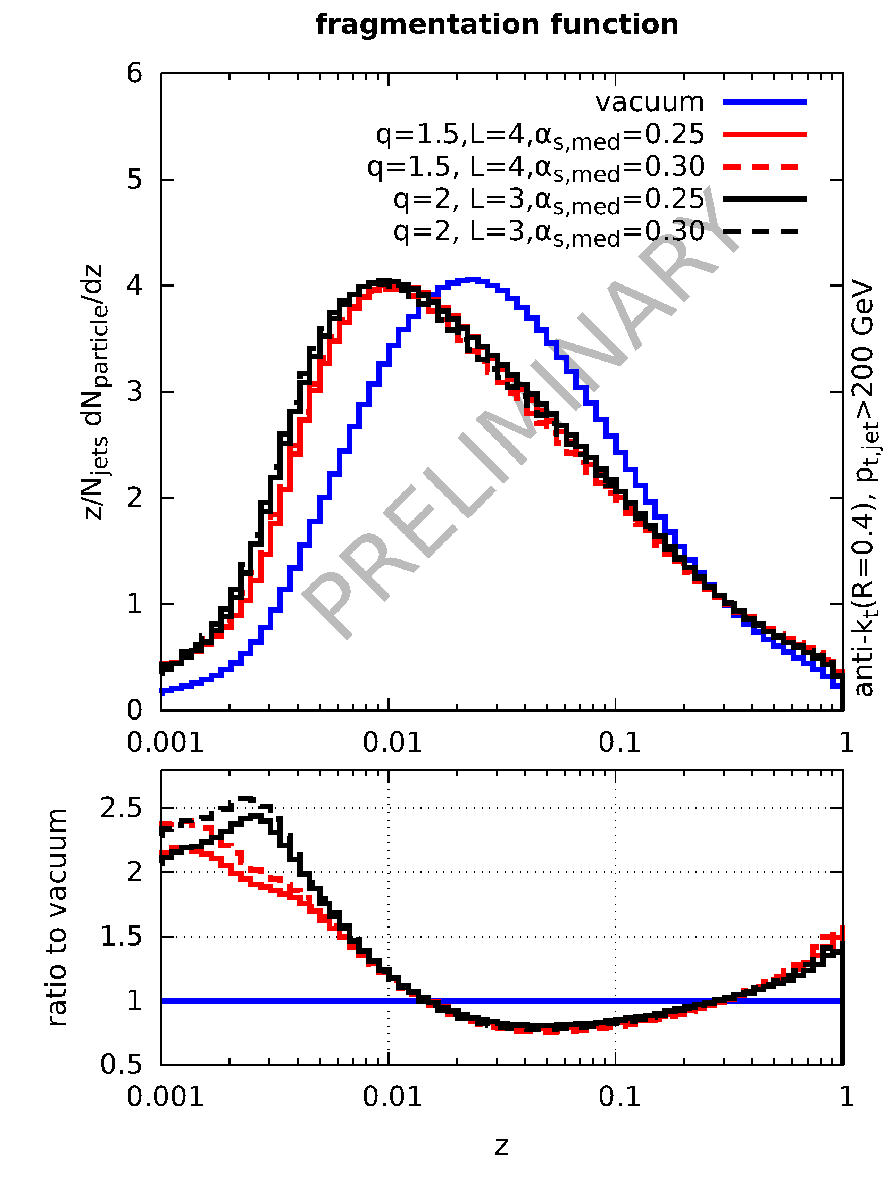}
  \end{tabular}
 \end{center}
\caption{ \label{plot5TeV}(Left) Jet $p_T$ spectrum in the vacuum and with the medium from our MC event generator for several values of $\hat{q}$ and $L$ and nuclear modification factor for jets $R_{AA}^{\ttiny{jet}}$. (Right) 
Fragmentation function at the partonic level. All jets with $p_{T,\ttiny{jet}}\ge200$ GeV have been selected for these analysis.}
\end{figure}

\subsection{Results and discussion}

In this section, some preliminary results of the full MC are shortly discussed. The hard process is generated from a list of hard events in di-jet production at LHC with 
$\sqrt{s}=5.02$ TeV. Then, the two partons created by this hard process are showered until the border of the in-medium region with the vacuum-like shower module.
All the partons created by this first cascade become seeds of medium-induced cascades handled by the medium-induced parton shower. 
Finally, the resulting particles of the previous shower are evolved in the ``outside'' region until hadronisation with the vacuum shower module. 
To account for color decoherence, all the partons produced inside the medium (via either vacuum-like or medium induced emissions)
are allowed to radiate at any angle.

For the jet reconstruction in the final state, the sofware package FastJet is used \cite{Cacciari:2011ma}. Jets are reconstructed by the anti-$k_T$ algorithm \cite{Cacciari:2008gp}
with $R=0.4$.

In figure \ref{plot5TeV}, we show the results for the nuclear modification factor for jets $R_{AA}^{\ttiny{jet}}$ and the ratio  $R_{D(z)}$ of the fragmentation function in nucleus-nucleus 
collisions and proton-proton collisions. Concerning the $R_{AA}^{\ttiny{jet}}$ factor, our MC predicts very well the flatness of the data \cite{Khachatryan:2016jfl,Aaboud:2018twu} at high $p_T$. Physically, 
this is a consequence of the fact that energy loss is stronger for a jet than for a single parton and is increasing with the jet energy (due to enhanced radiations). 

The ratio $R_{D(z)}$ shows the same qualitative behaviour as LHC data \cite{Chatrchyan:2014ava,Aaboud:2018hpb} notably an enhancement at both very small and relatively large values of $z$, and a depletion in between. 
In our picture the enhancement at low $z$ is due to the violation of angular ordering by the first emission outside the medium. 
The enhancement at large $z$ comes from the fact that a single hard parton inside the jet loses less energy than the jet as a whole.

\section{Conclusion}

In this proceeding, we briefly discussed a new picture, emerging from pQCD for the evolution of a jet
propagating through a dense QCD plasma. We have furthermore presented the first results of 
a Monte Carlo event generator which is grounded in this picture. Our preliminary results favorably compare 
to the data. In the future, we plan to perform more systematic studies of phenomenology and also to extend 
our theoretical framework by including missing physical ingredients such as hadronisation and a more 
realistic geometry of the medium.

A more thorough justification of the validity of our picture beyond DLA together with more details on the MC implementation will be presented in a shortcoming publication.

\vspace{0.25cm}
\textbf{Acknowledgements.}  The work of P.C., E.I. and G.S. is supported in
part  by  the  Agence  Nationale  de  la  Recherche  project ANR-16-CE31-0019-01.
The  work  of  A.H.M.  is  supported in part by the U.S. Department of Energy Grant
\# DE-FG02-92ER40699.

\bibliographystyle{JHEP}
\bibliography{biblio}

\end{document}